\documentclass[preprint,aps,prb,floatfix,superscriptaddress]{revtex4-2}

\usepackage{amsmath,amssymb, mhchem}
\usepackage{graphicx}
\usepackage[dvipsnames]{xcolor}
\usepackage{dsfont}
\usepackage[T1]{fontenc}
\usepackage{hyperref}
\usepackage{svg}
\usepackage[normalem]{ulem}
\usepackage{fancyhdr} 

\fancypagestyle{titlepage}{%
  \fancyhf{} 
  \fancyhead[C]{\small Distribution Statement A. Approved for public release: distribution is unlimited.}
}

\fancypagestyle{plain}{%
  \fancyhf{}
  \fancyhead[C]{\small Distribution Statement A. Approved for public release: distribution is unlimited.}
  
}

\newcommand*{\rom}[1]{\expandafter\@slowromancap\romannumeral #1@}

\begin{document}

    \renewcommand{\figurename}{Figure}

	\title{Geometry-Controlled Polarization Photocurrents in Scalable PtSe$_{2}$ Infrared Pixels}

	\author{Eunice Y. Paik}
    \email{eunice.y.paik.civ@army.mil}
	\affiliation{DEVCOM Army Research Laboratory, 2800 Powder Mill Road, Adelphi, MD, 20783 USA}

	\author{Owen A. Vail}
	\affiliation{DEVCOM Army Research Laboratory, 2800 Powder Mill Road, Adelphi, MD, 20783 USA}
    \affiliation{MIT Institute for Soldier Nanotechnologies, 500 Technology Square, Cambridge, MA, 02139 USA}
	
    \author{Madaline R. Marland}
    \affiliation{DEVCOM Army Research Laboratory, 2800 Powder Mill Road, Adelphi, MD, 20783 USA}
    \affiliation{Johns Hopkins University Chemical and Biomolecular Engineering, 3400 North Charles Street Baltimore, MD,21218 USA}

    \author{William A. Beck}
	\affiliation{DEVCOM Army Research Laboratory, 2800 Powder Mill Road, Adelphi, MD, 20783 USA}

    \author{Antonio Llopis-Jepsen}
    \affiliation{DEVCOM Army Research Laboratory, 2800 Powder Mill Road, Adelphi, MD, 20783 USA}

	\author{Wendy L. Sarney}
    \affiliation{DEVCOM Army Research Laboratory, 2800 Powder Mill Road, Adelphi, MD, 20783 USA}

    \author{Jeffery H. Leach}
	\affiliation{DEVCOM C5ISR Center, 10221 S Burbeck Rd, Fort Belvoir, VA 22060 USA}

	\author{Stefan Heiserer}
	\affiliation{Institute of Physics \& Center for Integrated Sensor Systems (SENS), University of the Bundeswehr Munich, Werner-Heisenberg-Weg. 39, 85577 Neubiberg, Germany}
    
    \author{Nikolas Dominik}
	\affiliation{Institute of Physics \& Center for Integrated Sensor Systems (SENS), University of the Bundeswehr Munich, Werner-Heisenberg-Weg. 39, 85577 Neubiberg, Germany}

    \author{Cormac Ó Coileáin}
    \affiliation{Institute of Physics \& Center for Integrated Sensor Systems (SENS), University of the Bundeswehr Munich, Werner-Heisenberg-Weg. 39, 85577 Neubiberg, Germany}

	\author{Paul B. Seifert}
	\affiliation{Institute of Physics \& Center for Integrated Sensor Systems (SENS), University of the Bundeswehr Munich, Werner-Heisenberg-Weg. 39, 85577 Neubiberg, Germany}

	\author{Georg S. Duesberg}
	\affiliation{Institute of Physics \& Center for Integrated Sensor Systems (SENS), University of the Bundeswehr Munich, Werner-Heisenberg-Weg. 39, 85577 Neubiberg, Germany}

	\author{Blair C. Connelly}
	\affiliation{DEVCOM Army Research Laboratory, 2800 Powder Mill Road, Adelphi, MD, 20783 USA}

	\author{George J. de Coster}
    \email{george.j.decoster.civ@army.mil}
	\affiliation{DEVCOM Army Research Laboratory, 2800 Powder Mill Road, Adelphi, MD, 20783 USA}
    \affiliation{MIT Institute for Soldier Nanotechnologies, 500 Technology Square, Cambridge, MA, 02139 USA}
    \affiliation{Institute of Physics \& Center for Integrated Sensor Systems (SENS), University of the Bundeswehr Munich, Werner-Heisenberg-Weg. 39, 85577 Neubiberg, Germany}

	\date{\today}

	\begin{abstract}

		\noindent Polarization-sensitive photodetectors provide multi-dimensional optical information beyond the capabilities of traditional intensity-based detectors. The noble metal dichalcogenide \ce{PtSe2} presents unique opportunities for tunable polarization detection due to its strong spin-orbit coupling, material stability, customizable broadband polarization responses, and direct compatibility with a wide range of substrates for back-end-of-line silicon integration. In this work, we demonstrate room-temperature near-infrared to mid-wavelength infrared polarization photoresponses using scalable, as-grown, wide-area \ce{PtSe2} films, and we show that the measured polarization response is reshaped by device geometry. Finite-element current-flow simulations reproduce the observed spatial redistribution of the polarization-sensitive response and localize the transverse polarization response near the pixel center. This geometry-enabled separation allows wavelength-dependent laser spot scans to distinguish symmetry-allowed photocurrents from contact-proximate, dichroism-mediated photothermal contributions. At near-infrared wavelengths, we observe spatial response that is consistent with linear-dichroic photothermoelectric currents, whereas mid-wavelength infrared measurements reveal a helicity-dependent photothermal contribution under oblique illumination. These results identify pixel boundary engineering as both a design lever and a diagnostic tool for scalable \ce{PtSe2} polarization-sensitive infrared pixels.

	\end{abstract}


	\maketitle


	\section{Introduction \label{ch:intro}}

    \noindent On-chip infrared (IR) polarimetric imaging represents a rapidly emerging paradigm in optoelectronics, offering multi-dimensional optical readout of material composition, surface profiles, target orientation, and material interfaces that is limited in traditional intensity-based IR imaging systems~\cite{liu_deep_2020, rogalski_infrared_2003, li_satellite-derived_2013, tyo_review_2006}. However, conventional polarimetric architectures are hindered by physical and structural constraints. Standard bulk semiconductor IR photodetectors (e.g., $\mathrm{In}_x\mathrm{Ga}_{1-x}\mathrm{Sb}$ or $\mathrm{Hg}_x\mathrm{Cd}_{1-x}\mathrm{Te}$) often require cryogenic cooling to suppress dark currents and lack intrinsic polarization sensitivity, necessitating the downstream integration of bulky rotating polarizer wheels~\cite{sabatke_optimization_2000, tyo_review_2006} or multi-pixel structures with metasurfaces~\cite{balthasar_metasurface_2017, arbabi_full-Stokes_2018, rubin_matrix_2019, li_monolithic_2020}.

    Nonlinear optical (NLO) responses in quantum materials offer a promising route for broadband, ultrafast, room-temperature IR polarization detection. Materials that feature broken inversion symmetry and strong intrinsic spin-orbit coupling (SOC) possess a Rashba-split band structure and optical spin angular momentum-based selection rules that elicit directional NLO photocurrent generation for incident light with circular or linear polarizations~\cite{ma_direct_2017, dhara_voltage-tunable_2015, ogawa_photocontrol_2014, sharifpour_enhanced_2026}. Specifically, circular and linear photogalvanic effects (CPGE and LPGE) enable the conversion of circular or linear polarization directly into a transverse or longitudinal DC current without requiring an applied bias~\cite{sipe_second-order_2000, kastl_ultrafast_2015, plank_review_2018}. In principle, these NLO effects offer a broad spectral range from the visible to the far-infrared, ultrafast intrinsic response ($\sim$100~fs – 10~ps)~\cite{mcIver_control_2012}, room-temperature operation, and direct electronic readout of optical helicity --- attributes that are not intrinsic to conventional bulk semiconductor IR detectors.

	Despite their promise, prior studies on quantum material-based polarimetry have suffered from a lack of scalability. For example, the topological insulator $\mathrm{Bi}_2\mathrm{Se}_3$ possesses spin-polarized Dirac surface states that host robust CPGE photocurrents~\cite{mcIver_control_2012}. However, high-quality $\mathrm{Bi}_2\mathrm{Se}_3$ synthesis requires ideal molecular beam epitaxial growth with specialized off-cut substrates~\cite{connelly_emergence_2024}, and the topological surface state suffers from degradation~\cite{kong_rapid_2011, benia_reactive_2011, yashina_negligible_2013}. Similarly, Weyl semimetals such as $\mathrm{TaAs}$ exhibit outstanding polarization-sensitive responsivities~\cite{ma_direct_2017}, but their fabrication relies on manual mechanical exfoliation or single crystal growth followed by focused ion beam (FIB) milling and transfer to the target substrates, limiting their scalability.      

	Transition metal dichalcogenides (TMDCs) have emerged as a highly versatile class of materials for scalable optoelectronics due to their strong spin-orbit coupling (on the order of 100~meV~\cite{xiao_coupled_2012, zhang_two-dimensional_2014}), high carrier mobility~\cite{zhao_high-electron-mobility_2017}, and robustness. Among them, noble metal dichalcogenides (NMDCs) — a distinct TMDC subclass containing group-10 metals — have generated significant interest due to their giant SOC~\cite{zhao_high-electron-mobility_2017, feng_giant_2025}, Dirac type-II topological states~\cite{huang_type-II_2016, zhang_experimental_2017}, and environmental stability~\cite{li_topological_2017}. Specifically, \ce{PtSe2} stands out as a layered van der Waals NMDC that exhibits a thickness-dependent transition from a semiconducting monolayer to a semi-metallic bulk state~\cite{wang_platinum_2019, li_layer-dependent_2021, heiserer_impact_2025}. This semi-metallic phase features a spin-orbit-coupled bulk band structure that enables continuous, broadband optical transitions extending down to the infrared and terahertz regimes~\cite{yim_high-performance_2016, gerlei_strain_2026, sefidmooye_long-wave_2021}. Crucially, \ce{PtSe2} exhibits high ambient stability and can be synthesized over wafer-scale areas at low temperatures ($\leq 450^{\circ}$C) via thermally assisted conversion (TAC) of pre-deposited platinum films~\cite{yim_high-performance_2016, boland_robust_2019, ansari_quantum_2019}. This synthesis temperature falls well within the thermal budget for back-end-of-line (BEOL) complementary metal-oxide-semiconductor (CMOS) processing, enabling direct, transfer-free integration of \ce{PtSe2} onto standard silicon photonic and electronic wafers. While prior time-domain terahertz spectroscopy studies on \ce{PtSe2} suggest the presence of NLO CPGE through substrate-induced inversion symmetry breaking~\cite{hemmat_layer-controlled_2023}, their use in polarization-sensitive photodetectors and sensors~\cite{parhizkar_two-dimensional_2022, wang_layered_2021} remains largely unexplored.

    In practical polarization-sensitive pixels, the collected photocurrent need not directly reflect the local optical generation mechanism. The measured response can be reshaped by finite conductivity, device boundaries, contact placement, and photothermal gradients. Consequently, a polarization-dependent signal measured at the contacts may contain both symmetry-allowed optical photocurrents and extrinsic contributions introduced by current collection and local absorptance. Establishing how pixel geometry affects these contributions is therefore necessary both for device design and reliable interpretation of polarization photocurrent measurements.  

    In this study, we use pixel geometry as a controlled electrical boundary condition for polarization-sensitive photocurrents in scalable, thermally assisted conversion (TAC)-grown \ce{PtSe2}. We combine symmetry analysis, finite-element current-flow modeling, and spatially resolved photocurrent measurements to determine how square, cloverleaf, and bow-tie pixels redistribute optically generated currents. We find that the conventional square geometry mixes nominally transverse and longitudinal current channels through boundary-constrained Ohmic return currents. Cloverleaf and bow-tie geometries reduce this mixing and produce a more centrally localized transverse polarization response. This geometry-enabled separation then allows wavelength-dependent spot scans at 976, 1550, and 4600 nm to identify distinct contact-proximate photothermal contributions to the measured polarization photocurrent. Our aim is to establish how pixel geometry reshapes the measured response and to use that understanding to separate intrinsic and photothermal contributions.

	\section{Results} \label{ch:results}

	\subsection{Symmetry-allowed polarization photocurrent channels in nanocrystalline \ce{PtSe2}}
	\begin{figure}[t]
		\centering
		\includegraphics[width=1\linewidth]{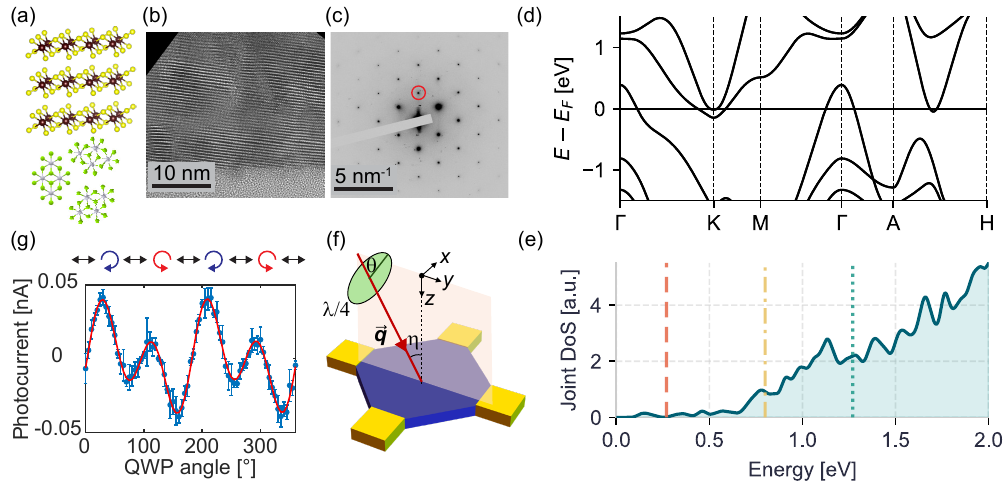} 
		\caption{Nanocrystalline \ce{PtSe2} device and experimental geometries. (a) Top: side-view schematic of 1T tri-atomic layered \ce{PtSe2}. Bottom: top-down view representation of the rotational disorders in nanocrystalline \ce{PtSe2}. (b) Transmission electron microscope (TEM) image of an approximately 36-nm thick nanocrystalline \ce{PtSe2} sample. (c) TEM electron diffraction pattern. The red circle denotes the diffraction spots corresponding to the (001) \ce{PtSe2} planes (upper right small spot) and the (002) Si planes (strong spot centered in circle).   (d) Calculated DFT band structure of bulk \ce{PtSe2}. (e) JDOS for direct optical transitions calculated using the bulk \ce{PtSe2} band structure. Vertical lines indicate excitation energies for optical measurements, corresponding to 4600 nm (dashed), 1550 nm (dot-dashed) and 976 nm (dotted). (f) Schematic representation of the \ce{PtSe2} pixel device and the experimental geometry. Laser with a wave vector $\vec{\textbf{q}}$ is incident the pixel along $y-z$ plane and at an angle $\eta$ from normal. The light passes through a quarter waveplate ($\lambda/4$) set to angle $\theta$ from the P-polarization of the beam. (g) A typical photocurrent $I(\theta$) versus quarter waveplate angle $\theta$ relationship from Equation~\ref{eq:I}.}
		\label{fig:main-intro}
	\end{figure}

	\noindent  The polarization-dependent photocurrent, $j_{i = x,y,z}$, in \ce{PtSe2} is a second-order NLO response to an oscillating electric field, $E_{i}(\omega)$, at frequency $\omega$ and wavenumber $\textbf{q}$. In the DC limit where $\omega \rightarrow 0$, it can be written as:
	\begin{equation} \label{eq:j}
		j_i = \int d\omega'\left[\sigma_{ijk}\left(\omega', -\omega'\right) + q_{l}T_{iljk}\left(\omega', -\omega'\right)\right] \cdot E_j(\omega')E_k^*(\omega'). 
	\end{equation}     
	\noindent $\sigma_{ijk}$ is a third rank photocurrent conductivity tensor that captures the photogalvanic effect (PGE), and $T_{iljk}$ is a fourth rank photon drag effect (PDE) tensor. The TAC grown \ce{PtSe2} is nanocrystalline and consists of many 1T tri-atomic domains rotated primarily about the out-of-plane axis as shown in Figure~\ref{fig:main-intro}(a). The transmission electron microscope (TEM) image in Figure~\ref{fig:main-intro}(b) shows that there are multiple 1T grains of various sizes. Since \ce{PtSe2} is deposited onto amorphous SiO$_{2}$, crystalline growth preferentially orients itself to its lowest-energy (001) basal planes parallel to the substrate to minimize surface energy. Similar morphology occurs in Bi$_{2}$Se$_{3}$ grown on SiO$_{2}$~\cite{jerng_ordered_2013}. The randomly in-plane oriented individual islands grow outward and eventually meet to form a continuous film. When the grain edges are pushed against each other, localized tilting of the (001) growth planes occurs. X-ray diffraction (XRD) (Figure~\ref{fig:SM-XRD}) further shows a dominant peak corresponding to the (001) plane, and a smaller peak for the (110) plane. 

    The TEM diffraction pattern Figure~\ref{fig:main-intro}(c) was collected with a 150 nm-selected area aperture; therefore, it includes contributions from a larger area that exceeds the region shown in the TEM image. Discrete spots, rather than continuous rings, provide evidence that the nanocrystals are not randomly oriented out-of-plane but instead share a limited number of crystallographic orientations. The strongest diffraction arises from the (001) \ce{PtSe2} planes, of which a majority are oriented $\sim$7$^{\circ}$ off from the Si (001) planes. The Moir\'e fringes indicate the in-plane polycrystallinity along the direction of the electron beam. Thus, averaged over the illuminated region, the nanocrystalline \ce{PtSe2} film is described by the point group $C_{\infty v}$: continuous rotational symmetry about the out-of-plane $z$-axis and in-plane mirror symmetry, with inversion symmetry broken by the substrate interface (Supporting Information). With this, the NLO tensors $\sigma_{ijk}$ and $T_{iljk}$ have very few non-zero elements. Following standard group theory analysis, Equation~\ref{eq:j} reduces to~\cite{connelly_emergence_2024, boyd_nonlinear_2008}: 

	\begin{equation} \label{eq:jxyz}
		\begin{pmatrix}
			j_x \\ j_y \\ j_z
		\end{pmatrix} = 
		\begin{pmatrix}
			C \sin 2\theta + L_1 \sin 4\theta \\
			L^y_2 \cos 4\theta + D^{y} \\
			L^z_2 \cos 4\theta + D^z
		\end{pmatrix}.
	\end{equation}

    Equation~\ref{eq:jxyz} provides the symmetry-allowed baseline for local polarization photocurrent generation in the nanocrystalline film. The measured current in a patterned device can additionally include geometry-dependent Ohmic redistribution and photothermal contributions, which are separated below through spatial mapping~\cite{song_shockley-ramo_2014, seifert_quantized_2019}, finite-element modeling, and wavelength-dependent measurements.
    
	\subsection{Pixel boundary engineering}  
	\noindent Locally, photo-induced currents in the nanocrystalline \ce{PtSe2} are expected to follow the directional and polarization-dependent response in Equation~\ref{eq:jxyz}. 
	The semimetallic bulk \ce{PtSe2} band structure enables broadband optical transitions. To confirm this, we performed density functional theory (DFT) calculations of the bulk \ce{PtSe2} band structure (Figure~\ref{fig:main-intro}(d)) and the resulting joint density of states (JDOS), $N_J(\omega)$ for allowed optical transitions at energy $\hbar \omega$:

    \begin{equation}
        N_{J}(\omega) = \sum_{n\neq m, \mathbf{k}} f_m(\mathbf{k})(1-f_n(\mathbf{k})) \delta(E_n(\mathbf{k})-E_m(\mathbf{k}) - \hbar \omega)~.
    \end{equation}

    \noindent Here $E_n(\mathbf{k})$ are the energy levels, and the Fermi-Dirac distribution terms, $f_n(\mathbf{k})$ accounts for Pauli blocking of transitions. While additional momentum and band dependent weight factors must be included to formally compute the shift and injection current contributions to PGE and PDE, $N_J(\omega)$ provides insight into which transitions are allowed (and resonant)~\cite{xie_photon-drag_2025}. The calculated $N_J(\omega)$ in Figure~\ref{fig:main-intro}(e) shows the semimetal nature of bulk \ce{PtSe2} permits continuous optical transitions down to arbitrarily small energy differences that correspond to long wavelength excitations, and reveals local features near 1.75 eV and 1.2 eV, which correlate with direct energy gaps between valence band maxima and conduction band minima.

	We then fabricated single-pixel devices and studied their optical responses under different excitation wavelengths. TAC-grown \ce{PtSe2} with 36-nm thickness was patterned into several detector configurations (Methods). To confirm the presence of polarization-dependent photocurrents, we optically excited the pixels with a laser incident at $\eta = 45^\circ$ using the experimental geometry shown in Figure~\ref{fig:main-intro}(f). The initial polarization of the incident laser was set along the y-axis, and was further controlled using a quarter waveplate (QWP) that produces circular polarization. The resulting photocurrent was collected along the $x$-direction, perpendicular to the plane of incidence ($j_{x,\perp}$), or along the $y$-direction, parallel to the plane of incidence ($j_{y,\parallel}$). A typical measured photocurrent $I(\theta)$ as a function of QWP angle $\theta$ is shown in Figure~\ref{fig:main-intro}(g) and takes the form:        

	\begin{equation} \label{eq:I}
		I(\theta) = D + C \sin 2\theta + L_1 \sin 4\theta + L_2 \cos 4\theta,
	\end{equation}
    
	\noindent where the coefficients represent the polarization independent ($D$), circular polarization ($C$) and linear polarization ($L_1, L_2$) dependent contributions to the total measured current.

	\begin{figure}[t]
		\centering
		\includegraphics[width=1\linewidth]{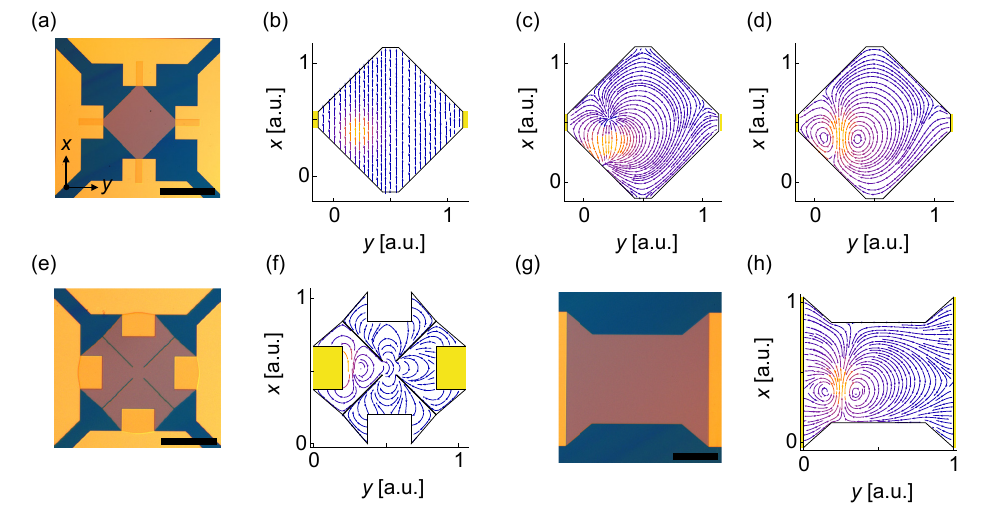}
		\caption{Optical microscope images and finite element simulated current flow plots of \ce{PtSe2} pixels. (a) Square pixel image. Black scale bars represent 250 $\mu$m. Simulated (b) optically-induced, (c) Ohmic, and (d) total current flow when the quarter waveplate is set to $\pi/8$ and the laser spot is off center. Yellow rectangles represent the connected electrical contacts. Cloverleaf (e) pixel image and (f) simulated total current flow. Bow-tie (g) pixel image and (h) simulated total current flow.}
		\label{fig:main-streamlines}
	\end{figure}

    To model the current flow and the boundary effects within the device, we turn to FE analysis. The current flow in the device was induced by incident light along the $y-z$ plane with a Gaussian beam profile chosen to match experimental conditions. Current was computed by solving the boundary value problem $\nabla \cdot \mathbf{J} = 0$, where $\mathbf{J}$ is the total current density $\mathbf{J} = \mathbf{J}_\text{opt} + \sigma\nabla V$, which is the sum of the optically-induced current density $\mathbf{J}_{\text{opt}}$ (Equation~\ref{eq:I}) and the Ohmic current $\sigma \nabla V$. The boundary conditions constrain the total current to flow parallel to the boundary and to $V =0$ at the connected contacts. The unconnected contacts are set to a potential that results in zero net current flow (Supporting Information). The potential gradient $\nabla V$ that builds to enforce this boundary condition bends the optically-induced current, mixing the $J_x$ and $J_y$ components. This is especially pronounced in the square geometry pixel (Figure~\ref{fig:main-streamlines}(a)). If the incident polarization of light is set such that optically induced current flows in the $x$-direction ($J_x$) (Figure~\ref{fig:main-streamlines}(b)), the addition of the Ohmic current (Figure~\ref{fig:main-streamlines}(c)) causes the total current (Figure~\ref{fig:main-streamlines}(d)) to swirl along the boundaries. Cutting notches to form cloverleaf geometry (Figures~\ref{fig:main-streamlines}(e, f)), or having a smaller center pixel region which fans out towards the contacts as in the bow-tie geometry (Figures~\ref{fig:main-streamlines}(g, h)) offers a way to control the direction of current flow and reduces the high-resistance obstructions to current flow. Current flow simulations for a centered optical spot are shown in Figure~\ref{fig:SM-streamlines}.

	\subsection{Photocurrent mapping of pixels}
	\begin{figure}[t]
		\centering
		\includegraphics[width=1\linewidth]{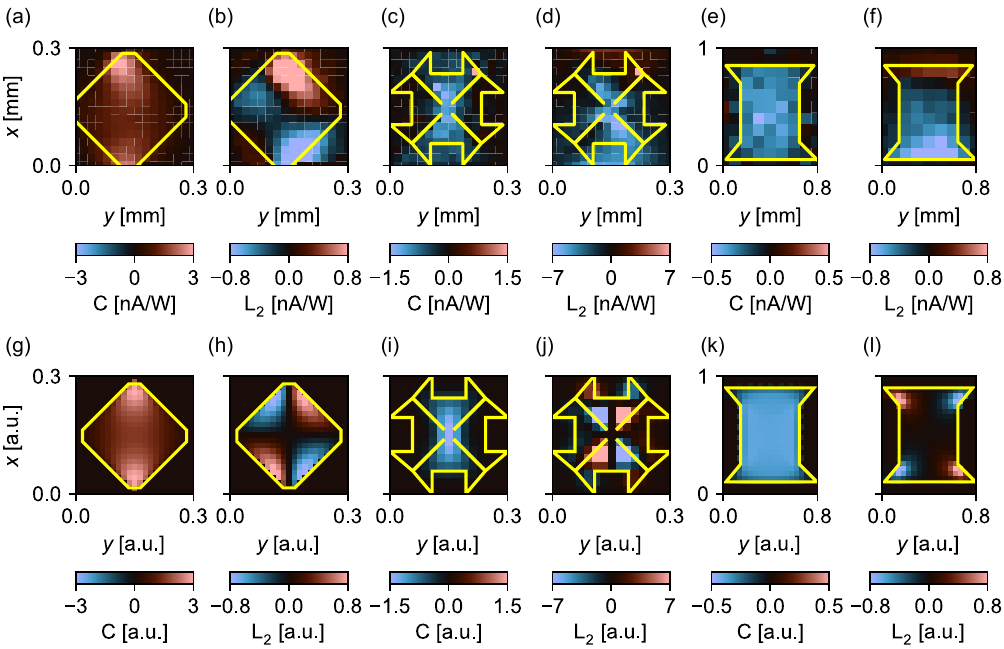}
		\caption{Position-dependent photocurrent harmonic coefficients $C$ and $L_2$ for current collected in the $j_{x, \perp}$ direction. (a - f) Experimental measurements and (g - l) finite element simulation of the harmonic coefficients. Laser plane-of-incidence is along the $y$-direction. Yellow solid line outlines the device.} 
		\label{fig:main-CL2map}
	\end{figure}

    To study the device geometries and their impact on the photocurrent distribution further, we performed 2D mapping of the photocurrent measurements of under-filled devices and extracted the NLO harmonic coefficients at each sample location by measuring the QWP dependence of the photocurrent for QWP angles from 0 to 360$^\circ$. Figure~\ref{fig:main-CL2map} shows the experimental (Figure~\ref{fig:main-CL2map}(a - f)) and finite-element simulated (Figure~\ref{fig:main-CL2map}(g - l)) results for the circular polarization, $C$, and $L_2$ linear polarization components for the current collected in the $j_{x,\perp}$ direction. The square device shows enhanced $C$ responses at the corners and weaker responses near the center (Figure~\ref{fig:main-CL2map}(a)), along with strong $L_2$ responses due to the current mixing between the $j_{x, \perp}$ and $j_{y, \parallel}$ channels (Figure~\ref{fig:main-CL2map}(b)). These spatial patterns are reproduced by the simulations (Figures~\ref{fig:main-CL2map}(g, h)). The cloverleaf and bow-tie device geometries mitigate the Ohmic back-flow currents such that the maximum $C$ response is at the center of the pixel, as shown in both experiment (Figure~\ref{fig:main-CL2map}(c, e)) and simulation (Figure~\ref{fig:main-CL2map}(i, k)). The $L_1$ photocurrent in the $j_{x,\perp}$ direction has the same spatial-dependence as $C$ (Equation~\ref{eq:jxyz}), and is shown in Figure~\ref{fig:SM-L1perp}. When the current is collected along the $j_{y, \parallel}$ direction, the $L_2$ component is dominant, as shown in Figure~\ref{fig:SM-para} and in agreement with Equation~\ref{eq:jxyz}.

	The offset current component, $D$, for all three devices is shown in Figure~\ref{fig:SM-D}. Based on the simplified behavior of intrinsic photocurrent response in disordered nanocrystalline \ce{PtSe2}, we expect $D = 0$ for $j_x$ and $D = 3L_2$ for $j_y$ (Supporting Information). In reality, photothermal effects near the contacts and edges give rise to finite $D$ in $j_x$ and $D > 3L_2$ in $j_y$, which emerge once the Ohmic backflow currents are suppressed. Figure~\ref{fig:SM-D} compares the $j_{x, \perp}$ and $j_{y, \parallel}$ components of $D$ and $D - 3L_2$, respectively, to show the non-intrinsic photothermal responses within the devices. These photothermal responses are also evident in the deviation of the experimental (Figures~\ref{fig:main-CL2map}(d, f)) results from the simulated (Figures~\ref{fig:main-CL2map}(j, l)) results for coefficient $L_2$ in the photocurrent mapping, where there are extrinsic polarization-dependent components that arise in the experimental data that are not fully captured with group theory analysis.

	\begin{figure}[t]
		\centering
		\includegraphics[width=1\linewidth]{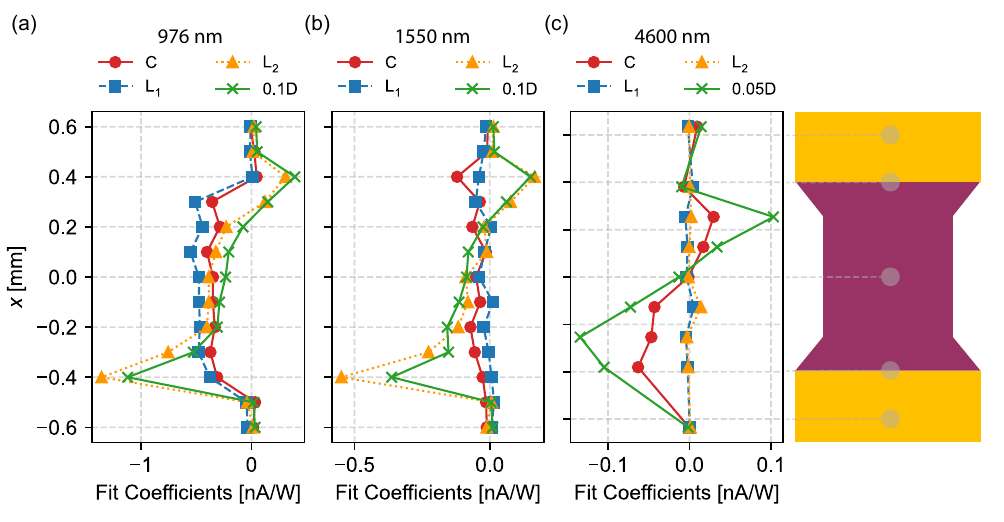}
		\caption{Wavelength-dependent harmonic coefficients along the length of the bow-tie device at (a) 976 nm, (b) 1550 nm, and (c) 4600 nm incident light at $\eta = 45^\circ$. Dashed lines connect the data points to the locations on the device schematic. The measured photocurrent is collected along the $j_{x, \perp}$ direction. For display, $D$ is scaled by 0.1 in (a, b) and by 0.05 in (c).} 
		\label{fig:main-wavedep}
	\end{figure}

	\begin{figure}[t]
		\centering
		\includegraphics[width=1\linewidth]{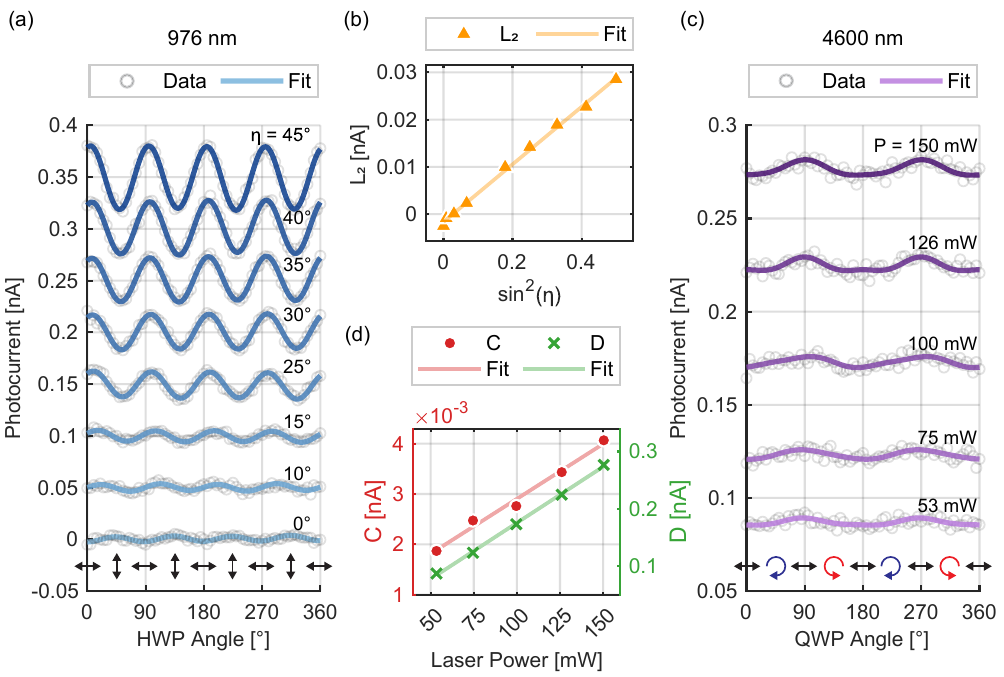}
		\caption{Dichroism-induced photothermal effect. (a) Linear polarization-dependent photocurrent as a function of angle of incidence, $\eta$, at 976-nm excitation. Artificial offset is added for readability. Solid lines are a fit to the functional form $L_2\cos{(4\psi + \phi)}$, where $\psi$ is the half-waveplate angle. The $\cos(4\psi)$ amplitude, $L_2$, is fitted and plotted as a function of the projection of the incident electric field vector in the out-of-plane direction in (b). (c) Photocurrent versus quarter-waveplate angle at several laser powers, at 4600-nm excitation. Solid lines represent a fit to the functional form $C\sin{(2\theta + \phi)} + D$. The circular photocurrent component ($C$) and offset current ($D$) are plotted as a function of laser power in (d).} 
		\label{fig:main-dichro}
	\end{figure}

	\subsection{Wavelength dependence and photothermal effects}
	\noindent To determine the source of the observed photothermal effects, we investigate the wavelength-dependent responses that probe different transitions in the band structure. As shown in Figures~\ref{fig:main-intro}(d) and (e), NIR optical transitions occur near the $\Gamma$ point, whereas the MWIR occur near the $K$ point. Line measurements were performed at excitation wavelengths of 976 nm, 1550 nm, and 4600 nm by scanning the laser spot along the $x$-direction of a bow-tie device. The excitation laser was incident in the $y-z$ plane at $\eta = 45^{\circ}$ and the current was collected along the $x$-direction ($j_{x, \perp}$) as a function of QWP angle to extract the harmonic coefficients and offset currents (Figure~\ref{fig:main-wavedep}).  Under 976-nm excitation, as shown in Figure~\ref{fig:main-wavedep}(a), the circular and first linear coefficients $C$ and $L_1$ maintain their maximum value through the center of the device and disappear as the laser is swept on top of the contacts. A finite offset current $D$ and $L_2$ linear coefficient can be seen to persist through the sample but change sign from top to bottom and vanish in the center of the device. This suggests a linear polarization-dependent photothermal effect. A similar behavior can be seen for measurements using 1550-nm excitation wavelength in Figure~\ref{fig:main-wavedep}(b), but the $C$ and $L_1$ responses are weaker at this wavelength. This is potentially due to the lack of a resonant transition at the corresponding photon energy 800 meV, as can be seen in the JDOS in Figure~\ref{fig:main-intro}(e). For 4600-nm (270 meV) excitation, a dominant helicity-dependent response is seen near the contacts, as evidenced by the predominantly two-fold periodicity in the photocurrent data and a strong $C$ coefficient (Figure~\ref{fig:main-wavedep}(c)). Notably, the circular coefficient $C$ changes sign between responses at the top and bottom of the device, closely mirroring the trend of the $D$ coefficient and suggesting a circular polarization-dependent photothermal effect. This behavior was reproduced in repeated measurements on separate days.  

    Generating a polarization-dependent photocurrent contribution from the photothermoelectric effect (PTE) necessitates optical dichroism within the material (Supporting Information Section~\ref{SM:PTE}). Under 976-nm excitation, the matching $L_2$ and $D$ behavior suggests linear dichroism. To further support this, we collected data with the laser spot aligned near the contact edge and varied the incident linear polarization by changing the half waveplate (HWP) angle, $\psi$. Interestingly, at $\eta = 45^\circ$, the photocurrent has a pronounced cos(4$\psi$) dependence, regardless of the relative orientation of the contact edge (Figure~\ref{fig:SM-HWPdep}). This indicates a preferential absorption of the P-polarized light due to the out-of-plane electric field component. To confirm this, we performed an angle-of-incidence-dependent photocurrent measurement, as shown in Figure~\ref{fig:main-dichro}(a). The absorption associated with the out-of-plane electric field vector scales as $\propto E^2_z \propto \sin^2{(\eta)}$. Plotting the $L_2$ (or cos(4$\psi$) component) as a function of $\sin^2{(\eta)}$, we see a linear dependence (Figure~\ref{fig:main-dichro}(b)), indicating a preferential absorption of the P-polarized light and an oblique angle-of-incidence-induced linear dichroism. For a type-II Dirac semimetal such as \ce{PtSe2}, the tilt of the Dirac cone leads to an out-of-plane optical conductivity $\sigma_{zz}(\omega)$, contributing to both intra- and interband transitions~\cite{li_anisotropic_2024}. In the NIR, interband transitions occur at the tilted Dirac cone near the $\Gamma$ point, leading to a preferential absorption of P-polarized light~\cite{zhang_experimental_2017}. This linear-dichroic PTE at 976-nm and 1550-nm excitation likely accounts for the discrepancy between the experimental (Figures~\ref{fig:main-CL2map}(d, f)) and simulated (Figures~\ref{fig:main-CL2map}(j, l)) $L_2$ coefficients for the cloverleaf and bow-tie devices that have relatively larger contact areas. Adding a phenomenological, contact-localized thermal profile to the FE simulation qualitatively reproduces the experimental $L_2$ maps (Figure~\ref{fig:SM-thermalsim}).   
    
    Under 4600-nm excitation, the matching $C$ and $D$ behavior suggests the presence of circular dichroism. Figure~\ref{fig:main-dichro}(c) shows a power-dependence of the photocurrent. Figure~\ref{fig:main-dichro}(d) shows that the circular ($C$) and offset ($D$) components both scale linearly with laser power. Because linear scaling is expected for both intrinsic photogalvanic and photothermal currents, the power dependence alone does not discriminate between them; the principal evidence for a circular-dichroic PTE is the spatial sign reversal of $C$ that tracks $D$ across the device (Figure~\ref{fig:main-wavedep}(c); Supporting Information Section~\ref{SM:PTE}). The 4600-nm excitation wavelength corresponds to a photon energy of $\hbar \omega = 0.27$ eV, where the calculated interband JDOS (Figure~\ref{fig:main-intro}(e)) is small, and corresponds to excitations near the K point of \ce{PtSe2}, typically probed with longer wavelength excitations~\cite{li_topological_2017, gerlei_strain_2026}. We can therefore expect Fermi surface excitations and the PTE to be more relevant to photocurrent generation than interband NLO transitions. Previous work has shown that polycrystalline TAC \ce{PtSe2} deposited on a substrate with similar thicknesses to the films studied here exhibit circular dichroism~\cite{hemmat_layer-controlled_2023}. Corresponding DFT calculations in that work showed that by breaking inversion symmetry in \ce{PtSe2} through substrate effects, typical TMDC valley circular polarization selection rules emerge at the K and K$'$ points~\cite{sallen_robust_2012}. However, valley selection rules alone should not lead to an overall net preferential absorptance of left vs. right circular polarization, as the K and K$'$ will preferentially absorb the opposite helicities equally, i.e. $\alpha_{K}(\text{LCP}) = \alpha_{K'}(\text{RCP})$. TEM diffraction and XRD measurements indicate that the \ce{PtSe2} film is locally polycrystalline and largely random in-plane, but with a small net crystallite canting relative to the substrate normal, which can provide a nonzero average tilt vector over the illuminated region (see Figure~\ref{fig:main-intro} and Figure~\ref{fig:SM-XRD}). Under excitations at an oblique incidence, this texture allows an extrinsic circular-dichroic contribution to the absorptance, which would then contribute a helicity modulation to the PTE generated photocurrent (see Supporting Information).

	\section{Conclusion} \label{ch:conclusion}

	\noindent In summary, we have demonstrated that scalable, large-area \ce{PtSe2} synthesized via low-temperature TAC is a promising material platform for room-temperature, broadband NIR-to-MWIR polarization-sensitive photodetection. By fabricating devices directly on nanocrystalline films, we have bypassed the constraints of single-crystal alignment present in topological insulators and Weyl semimetals. Group theory analysis separates the circular ($C$) and linear ($L_1$) polarization terms into the transverse photocurrent channel ($j_{x, \perp}$) while the orthogonal linear ($L_2$) polarization and the polarization-independent ($D$) component contribute to the longitudinal photocurrent channel ($j_{y, \parallel}$). Replacing the square pixel with cloverleaf and bow-tie geometries reduces boundary-induced mixing between these channels and shifts the maximum circular response from the pixel corners to the pixel center, in agreement with finite-element current-flow simulations.

    Because geometry suppresses Ohmic mixing, wavelength-dependent spot scans can separate intrinsic polarization photocurrents from extrinsic PTE contributions near the contacts. In the NIR, the contact-proximate response is dominated by a linear-dichroic PTE, evidenced by the $\sin^2\eta$ scaling of $L_2$ and its spatial correlation with $D$. In the MWIR, we observe a helicity-dependent PTE whose sign tracks $D$ across the device, which we attribute to a circular-dichroic absorptance arising from a small net crystallite canting under oblique illumination. Such dichroism-mediated channels could, in principle, supply additional linearly independent polarization responses toward single-pixel Stokes readout~\cite{jia_taas_2022, khajavikhan_weyl_2025}.

    Together, geometry-controlled current collection and dichroism-mediated photothermal channels suggest a path toward chip-scale, uncooled IR polarization-sensitive arrays, and the BEOL compatibility of TAC synthesis allows such pixels to be integrated directly onto readout circuits or silicon photonics. Quantitative optimization of pixel geometry against polarimetric figures of merit, together with control of film thickness and crystallite canting, is the subject of ongoing work and will be required for independent readout of the Stokes parameters.

	\begin{acknowledgments}

    This research was supported by a Laboratory University Collaborative Initiative award provided by the Basic Research Office in the Office of the Under Secretary of Defense for Research and Engineering and by the DEVCOM Army Research Office and was accomplished under Cooperative Agreement Number W911NF2520010 (STEP-TWO). The views and conclusions contained in this document are those of the authors and should not be interpreted as representing the official policies, either expressed or implied, of the Army Research Office or the U.S. Government. The U.S. Government is authorized to reproduce and distribute reprints for Government purposes notwithstanding any copyright notation herein. GJdC would like to acknowledge funding from the US Office of the Deputy Assistant Secretary of the Army for Defense Exports and Cooperation Engineering and Scientist Exchange Fellowship program. We thank dtec.bw—Digitalization and Technology Research Center of the Bundeswehr for support (project VITAL-SENSE). dtec.bw is funded via the German Recovery and Resilience Plan by the European Union (NextGenerationEU)

	\end{acknowledgments}

	\section{Methods}
    
    \subsection{Material Growth}
    \noindent \ce{PtSe2} thin films were synthesized using a standard thermally assisted conversion (TAC) method~\cite{ansari_quantum_2019, yim_high-performance_2016}. Platinum (Pt) was deposited onto n++-doped Si chips with a $\sim$330-nm SiO$_2$ layer with a Cressington 160 1SXD3 108auto sputter coater. Subsequently, the films were converted to \ce{PtSe2} at 450 $^\circ$C, in the presence of a selenium flux, produced by flowing 100 sccm H$_2$/N$_2$ mix across a Se source heated to 228 $^\circ$C.

    \subsection{Density functional theory (DFT) and JDOS calculations} 
    \noindent Density functional theory (DFT) calculations were performed using the Vienna Ab initio Simulation Package (VASP, version 6.4). Exchange-correlation effects were treated within the generalized gradient approximation (GGA) using the optB88-vdW functional of Klime\u{s} et al.~\cite{klimevs_chemical_2010}, which reproduces the \ce{PtSe2} experimental lattice parameters without sacrificing its band character description~\cite{villaos_thickness_2019}. Projector augmented wave (PAW) pseudopotentials with a plane-wave cutoff energy of 520 eV were used. Valence electrons were considered as Pt: $5d^96s^1$ and Se: $4s^2, 4p^4$. For the Brillouin zone sampling, Monkhorst-Pack k-point meshes of 20$\times$20$\times$10 and 36$\times$36$\times$18 were employed for structural relaxation and electronic structure calculations, respectively. The energy and force convergence criterion were set to $10^{-7}$ eV and $10^{-6}$ eV/{\AA}. Spin-orbit coupling (SOC) was included in all simulation steps. Following structural relaxation, the calculated lattice parameters were determined to be a = 3.80 {\AA} and c = 5.15 {\AA}, within 3\% of the scanning transmission electron microscope (STEM)-determined values (a = 3.70 {\AA} and c = 5.06 \AA~\cite{wang_monolayer_2015}). 


    \subsection{Device fabrication}
    \noindent Devices were defined using a Vistec EBPG5000 + ES electron beam lithography system (100 kV beam) and argon plasma etching (Oxford Plasma Lab 100) for variable contact/device geometry in a single run. Ti/Au (10/200 nm) metal contacts formed with consistent and symmetric rectangular shape reduced the expected impact of thermal effects within the gold contact itself. The design of the square pixel was inspired by Van der Pauw Hall measurements while the cloverleaf design intentionally overfilled the contact region. The sample was wire-bonded and mounted to a rotation stage for optical measurement.

	\subsection{Photocurrent measurements} 
    \noindent Optical characterization of \ce{PtSe2} pixels was performed primarily with a 976-nm (1.27 eV) continuous wave (cw) laser, which is near the 1.2 eV resonance shown in the JDOS (Figure~\ref{fig:main-intro}(e)). The incident light goes through a chopper and a set of polarization optics (half and quarter waveplates) to control the incident polarization. The light was then focused onto the device using an objective lens. The resulting photocurrent was collected via a lock-in amplifier referenced to the chopper frequency. 2D mapping measurement was done with motorized micromanipulator stages. The wavelength-dependent measurements were done with a 976-nm (1.27 eV), 1550-nm (0.8 eV), and 4600-nm (0.27 eV) cw lasers. The laser spot size is around 25 $\mu$m for the 976-nm and 1550-nm lasers and around 200 $\mu$m for the 4600-nm laser. The optics were mounted onto an optical breadboard rotation mount which can be rotated to change the angle of incidence. At 4600-nm, the narrow-band polarization optics can introduce uncertainty in the retardance of the prepared linear polarization, which we account for by introducing a phase term in the fits.

	\subsection{Finite element (FE) simulation}
	\noindent A FE mesh was prepared in \textit{Mathematica} for each corresponding device geometry. Along the device boundary, excluding intersections with the contact regions, a Neumann boundary condition (BC) was used to specify that no current flows perpendicular to the boundary. For each experimental geometry investigated, the corresponding contact pairs and optical response defined in (Equation~\ref{eq:jxyz}) are treated consistently by the simulations. In the contacts where the measurement is performed (i.e., the current is sampled), Dirichlet BCs were used to specify $V = 0$. For structures like square (Figure~\ref{fig:main-streamlines}(a)) and cloverleaf (Figure~\ref{fig:main-streamlines}(e)) that have two pairs of contacts, current flows through one contact pair and the other contact pair was left unconnected (open circuited). At the unconnected contacts, the impedance matrix and voltage values along the contacts were calculated such that no net current flows through the contact pair. The current was computed and integrated along the contact boundaries to determine the total current flowing through each contact. 

	A Gaussian-shaped intensity distribution of the optical excitation spot was used to generate optical current on the device. Resulting current was calculated at 24 equally spaced quarter waveplate angles, $\theta$, between 0 and $2\pi$, which were then fitted to extract the harmonic coefficients $C$, $L_1$, and $L_2$ at each location. The polarization dependence of $D$ is not included in the simulations. The spatial pattern and relative magnitudes of current are significant but the absolute magnitude is an arbitrary quantity determined by optical intensity and sheet conductivity $\sigma$.

    \subsection{Transmission Electron Microscopy (TEM)}
    \noindent Cross-sectional TEM samples were prepared using a Thermo Scientific Helios G4 UX Dual Beam focused ion beam (FIB) system. The samples were examined in a JEOL ARM 200F STEM operated at 200 keV with the beam direction coinciding along the [$\bar{1}$10] direction corresponding to the underlying Si substrate.

    \subsection{X-ray Diffraction (XRD)}
    \noindent Figure~\ref{fig:SM-XRD} shows x-ray diffraction (XRD) collected with a Panalytical X’Pert Pro. Figure~\ref{fig:SM-XRD}(a) is a coupled $2\theta-\omega$ scan that penetrates into the substrate and only includes planes that are aligned along the growth direction. Figure~\ref{fig:SM-XRD}(b) is a glancing incidence $2\theta-\omega$ scan that does not reach the substrate. 

    In Figure~\ref{fig:SM-XRD}(a) we see peaks corresponding to the (004) Si and the (001) and (110) \ce{PtSe2} planes. In Figure~\ref{fig:SM-XRD}(b) we removed the receiving slit and see planes corresponding to (001) and what is likely the (101) \ce{PtSe2} planes. The unlabeled peak does not correspond to any \ce{PtSe2} diffractions and could be due to elemental Se residing on the sample surface. The XRD scans are sampling a much larger area than that examined in the TEM.

    \bibliography{PtSe2_geometry_v3.bib}

    \clearpage
    
    \section{Supporting Information}

    \setcounter{figure}{0}
    \renewcommand{\figurename}{Figure}
    \renewcommand{\thefigure}{S\arabic{figure}}

    \begin{figure}[h]
		\centering
		\includegraphics[width=1\linewidth]{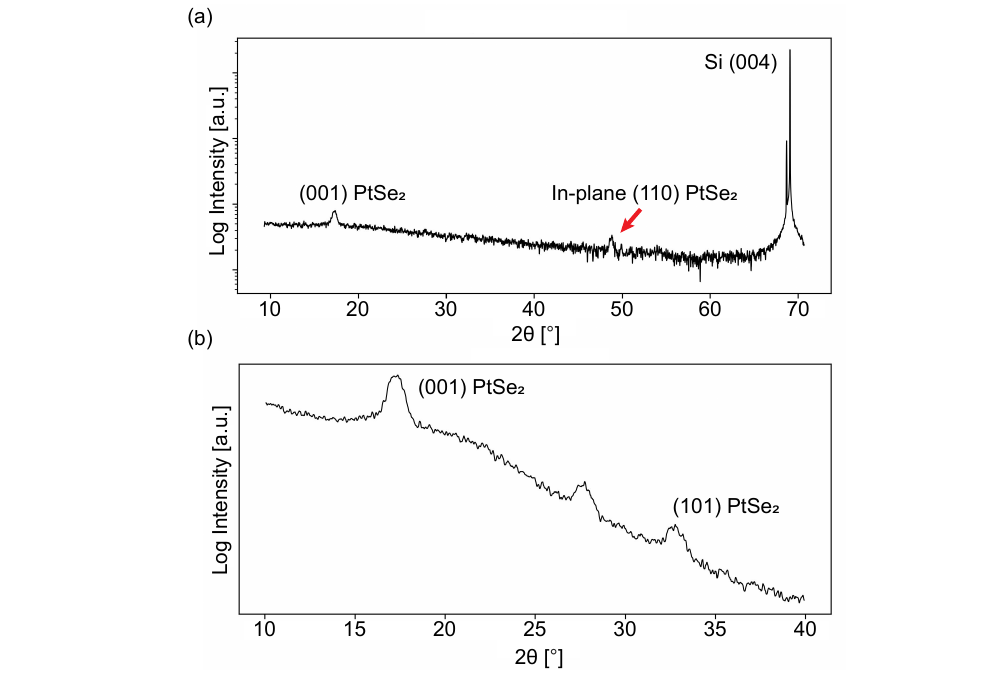}
		\caption{X-ray diffraction (XRD) of bulk \ce{PtSe2}. (a) A coupled $2\theta-\omega$ scan. (b) A glancing incidence $2\theta-\omega$ scan.} 
		\label{fig:SM-XRD}
	\end{figure}

    \begin{figure}[h]
		\centering
		\includegraphics[width=1\linewidth]{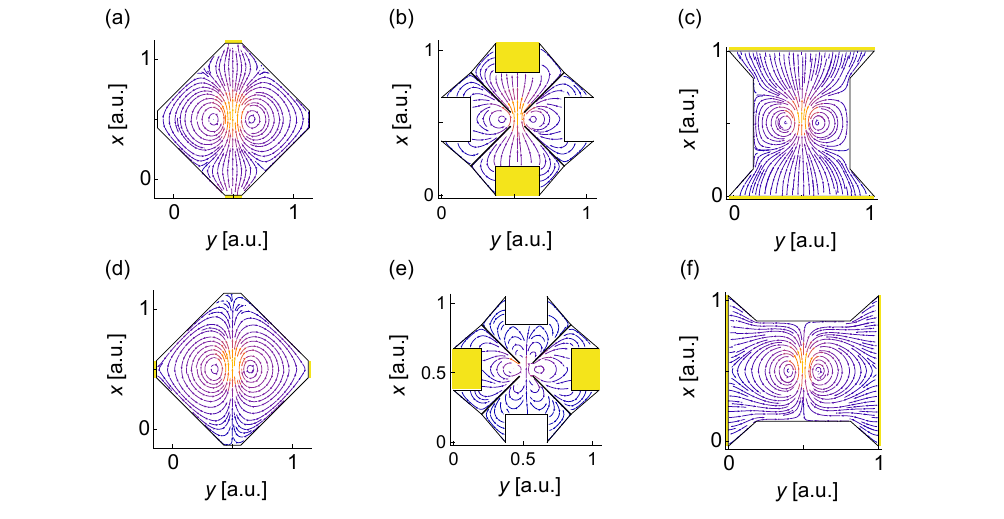}
		\caption{Simulated total current flow with a Gaussian optical spot that is centered on the device. (a - c) Current collected along the $j_{x, \perp}$ direction. (d - f) Current collected along the $j_{y, \parallel}$ direction.} 
		\label{fig:SM-streamlines}
	\end{figure}

    \begin{figure}[h]
		\centering
		\includegraphics[width=1\linewidth]{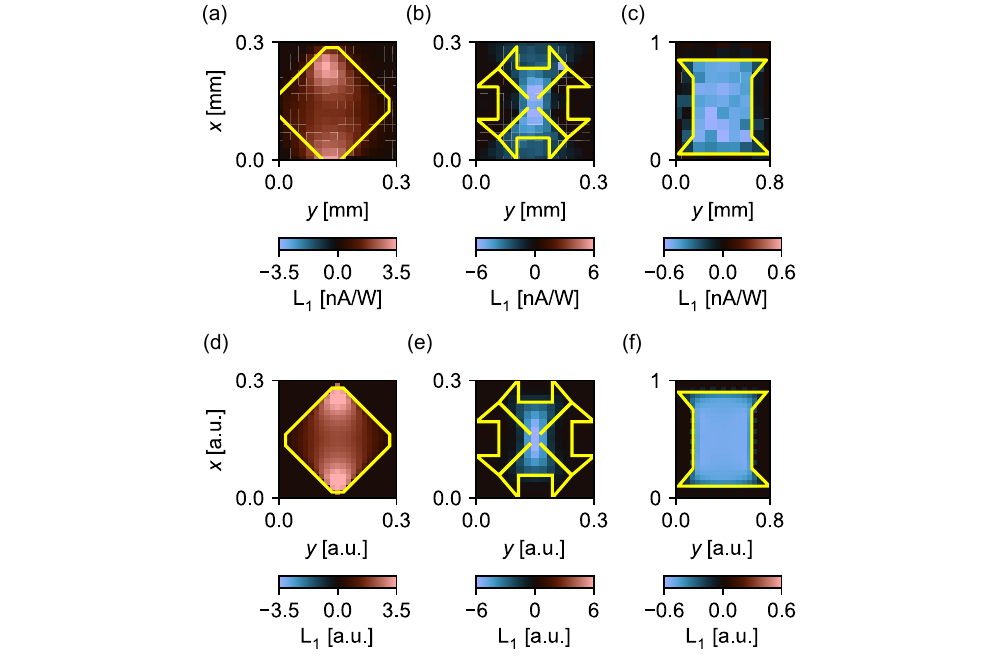}
		\caption{Photocurrent mapping of the $L_1$ coefficient. (a - c) Experimental and (d - f) finite simulation results.} 
		\label{fig:SM-L1perp}
	\end{figure}

    \begin{figure}[h]
		\centering
		\includegraphics[width=1\linewidth]{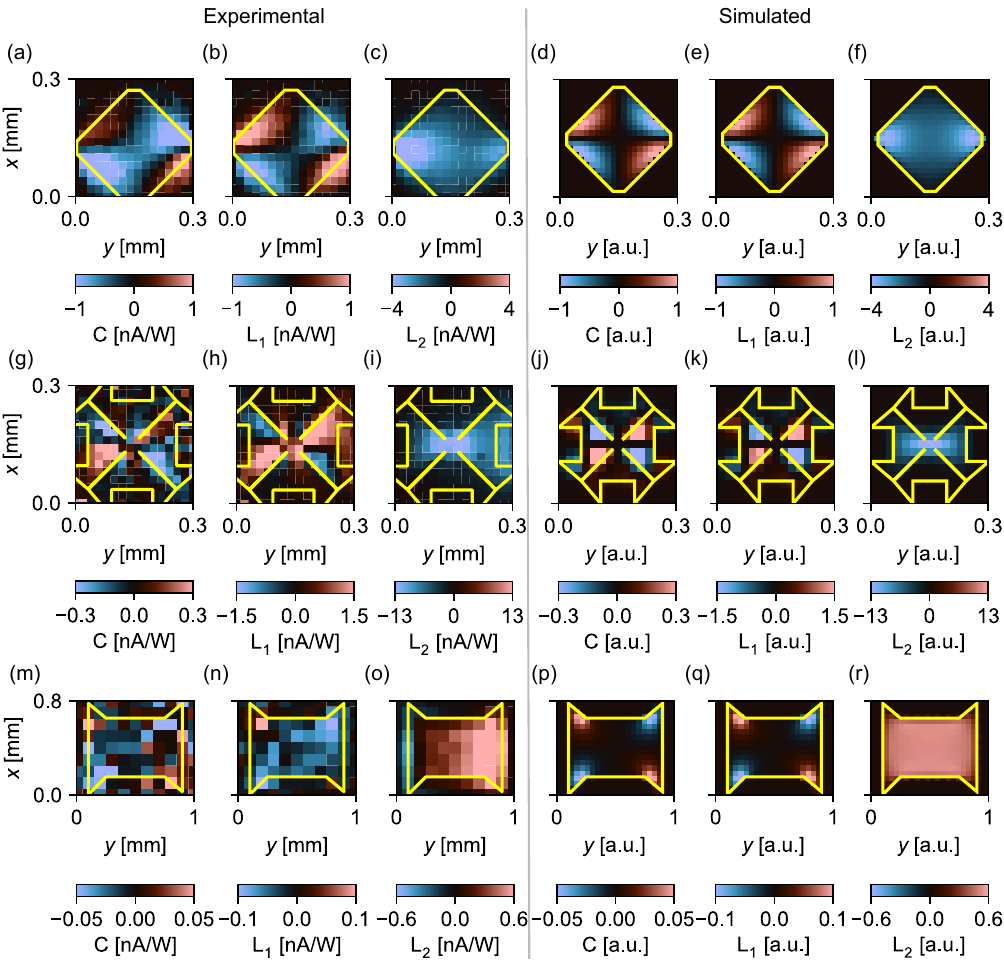}
		\caption{Position-dependent photocurrent harmonic coefficients for current collected along the $j_{y, \parallel}$ direction.} 
		\label{fig:SM-para}
	\end{figure}

    \begin{figure}[h]
		\centering
		\includegraphics[width=1\linewidth]{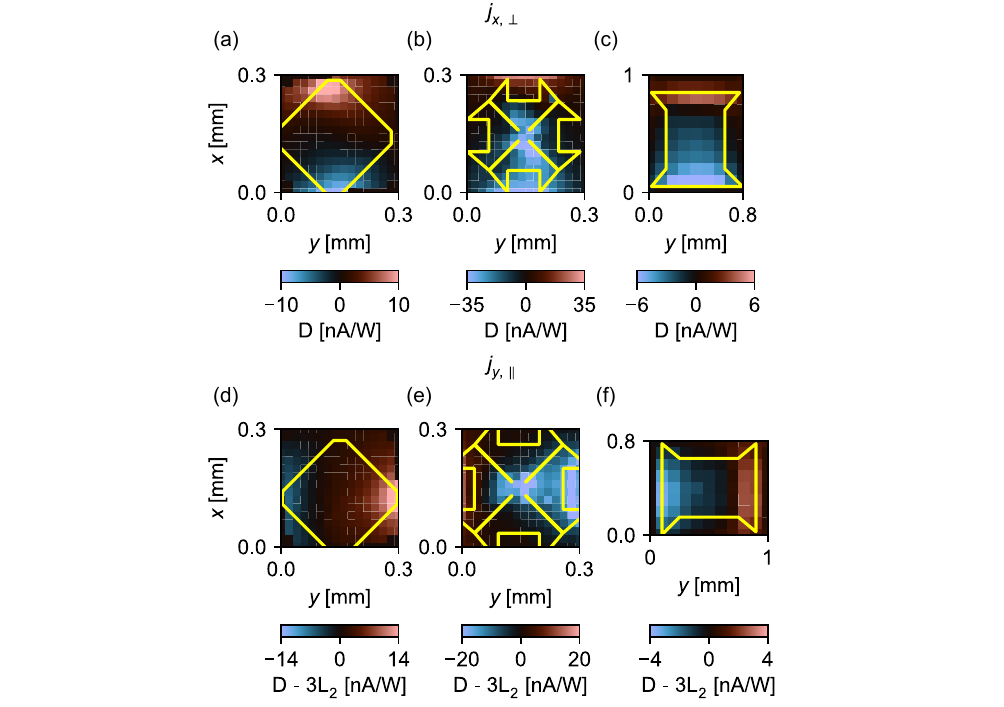}
		\caption{Photothermal component of the photocurrent. (a - c) shows the $D$ component of the $j_{x, \perp}$ photocurrent and (d - f) shows the $D - 3L_2$ component of the $j_{y, \parallel}$ photocurrent.} 
		\label{fig:SM-D}
	\end{figure}

    \begin{figure}[h]
		\centering
		\includegraphics[width=0.5\linewidth]{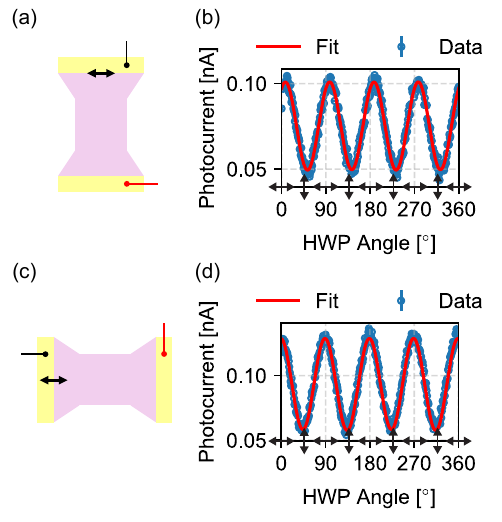}
		\caption{Linear polarization-dependent photocurrent at the bow-tie device contact edge, at $\eta = 45^\circ$. (a) Schematic of the bow-tie device at $j_{x, \perp}$ collection. Double-headed arrow indicates the location of the laser spot and the direction of horizontally polarized (P-polarized) light relative to the contacts. (b) Resulting photocurrent for the device orientation shown in (a). (c) Schematic of the bow-tie device at $j_{y, \parallel}$ collection, and resulting photocurrent shown in (d).} 
		\label{fig:SM-HWPdep}
	\end{figure}

    \begin{figure}[h]
		\centering
		\includegraphics[width=1\linewidth]{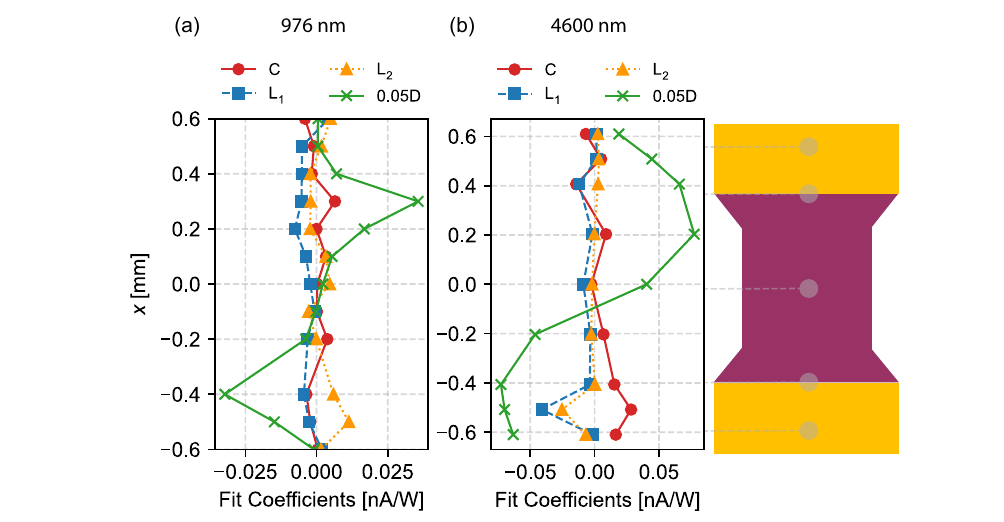}
		\caption{Wavelength-dependent harmonic coefficients along the bow-tie device at normal incidence ($\eta = 0$). (a) 976 nm and (b) 4600 nm excitation wavelengths. The measured photocurrent is collected along the $j_{x, \perp}$ direction.} 
		\label{fig:SM-wavedepnorm}
	\end{figure}

    \begin{figure}[h]
		\centering
		\includegraphics[width=0.5\linewidth]{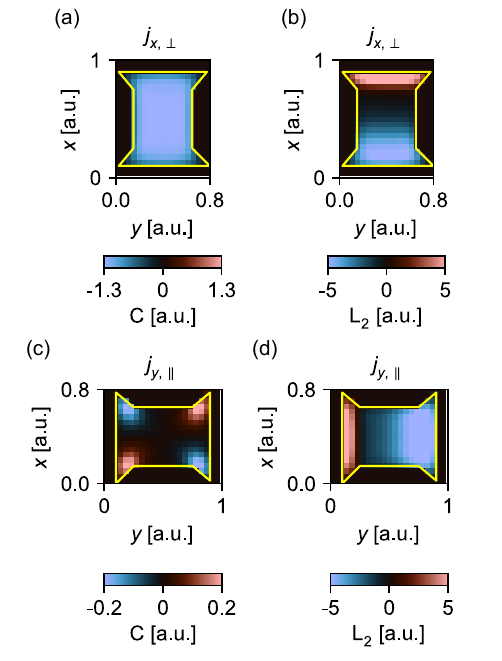}
		\caption{Finite element simulation of the photocurrent harmonic coefficients that includes a thermal photocurrent component. (a) $C$ and (b) $L_2$ components of the $j_{x, \perp}$ photocurrent. (c) $C$ and (d) $L_2$ components of the $j_{y, \parallel}$ photocurrent.} 
		\label{fig:SM-thermalsim}
	\end{figure}

    \subsection{Nonlinear Response Coefficients} \label{SM:nonlinear_coeffs}
    \noindent The surface symmetry group of crystalline \ce{PtSe2} is $C_{3v}$: three fold rotational symmetry with a $y-z$-mirror plane. The TAC \ce{PtSe2} is polycrystalline with random orientations, so we approximate the point group symmetry as $C_{\infty v}$, which is generated by arbitrary rotations about the $z$-axis $R_z(\varphi)$ and the $y-z$-plane mirror symmetry operator $M_{yz} = \text{diag}(-1,1,1)$. Following standard symmetry analysis of second order response tensors, we find:

    \begin{eqnarray}
    \sigma = 
    \begin{pmatrix}
     \left(\begin{smallmatrix} 0 \\ 0 \\ \sigma_{xxz} \end{smallmatrix}\right) &
     \left(\begin{smallmatrix} 0 \\ 0 \\ 0 \end{smallmatrix}\right) &
     \left(\begin{smallmatrix} \sigma_{xzx} \\ 0 \\ 0 \end{smallmatrix}\right) \\
     \\
     \left(\begin{smallmatrix} 0 \\ 0 \\ \sigma_{zxx} \end{smallmatrix}\right) &
     \left(\begin{smallmatrix} 0 \\ \sigma_{xxz} \\ 0 \end{smallmatrix}\right) &
     \left(\begin{smallmatrix} 0 \\ 0 \\ 0 \end{smallmatrix}\right) \\
     \\
     \left(\begin{smallmatrix} 0 \\ 0 \\ 0 \end{smallmatrix}\right) &
     \left(\begin{smallmatrix} 0 \\ \sigma_{zxx} \\ 0 \end{smallmatrix}\right) &
     \left(\begin{smallmatrix} 0 \\ 0 \\ \sigma_{zzz} \end{smallmatrix}\right)
    \end{pmatrix}
    \end{eqnarray}

    \noindent For the electric field incident on our sample described in Figure~\ref{fig:main-intro}, the resulting photocurrent $j_i = \sum_{jk} \sigma_{ijk} E_j E_k$ is found to be:

  \begin{eqnarray}
    \begin{pmatrix}
        j_x \\ j_y \\ j_z
    \end{pmatrix} 
    = E_0^2
    \begin{pmatrix}
     \sin(\eta) \left[ \frac{i}{2} \sin(2\theta) (\sigma_{xxz} - \sigma_{xzx}) - \frac{1}{4} \sin(4\theta) (\sigma_{xxz} + \sigma_{xzx}) \right] \\
     -\frac{1}{8}\sin(2\eta) (\sigma_{xxz} + \sigma_{xzx}) \left[ \cos(4\theta) + 3 \right] \\
     \frac{1}{4}\sin^2(\eta)     
     (\sigma_{zzz} - \sigma_{zxx}) \left[ \cos(4\theta) + 3 \right] + \frac{1}{2} \sigma_{zxx}
    \end{pmatrix}
    \end{eqnarray}  
    
    \noindent Note that $(\sigma_{xxz} - \sigma_{xzx})$ is purely imaginary for the circular response, so $C \propto i(\sigma_{xxz} - \sigma_{xzx})$ is real.

    \subsection{Photothermoelectric Currents and Dichroism} \label{SM:PTE}
    \noindent In the NIR and MWIR measurements, the polarization-independent photocurrent contribution, $D$, increased in strength as the laser spot was scanned away from the center of the device and approached device contacts (Figure~\ref{fig:main-wavedep} of the main text); the sign of $D$ reversed depending on which contact was nearest. This is the hallmark behavior of the photothermoelectric effect (PTE) contribution to the photocurrent in contacted devices~\cite{kastl_ultrafast_2015, dai_high-performance_2022, cai_giant_2025}, where asymmetric excitation in the device can lead to a thermal-gradient-induced current response. The PTE voltage is computed from the material Seebeck coefficient $S$ and thermal gradient, $\nabla T$, by:

    \begin{equation}
        V_{PH} = \int S(x) \partial_x T(x) dx~,
    \end{equation}

    \noindent where for simplicity we assume the gradient is only along the $x$-direction. Multi-layered \ce{PtSe2} has been observed to have a Seebeck coefficient $S>100\mu$V$/$K, which would enable a robust PTE at the \ce{PtSe2}/contact interface where the thermal gradient will be most pronounced. If we assume that the photocurrent generated by the PTE is proportional to $V_{PH}$, and that the temperature gradient is induced by laser heating, we can express the measured photocurrent $I_{\rm PTE}$ as a function of laser spot $x_0$ and polarization as:

    \begin{equation}
        I_{\rm PTE}(x_0,P)=K(x_0)\alpha(P)~,
    \end{equation}

    \noindent where $\alpha(P)$ is the absorptance for polarization $P$, and $K(x_0)$ comprises the laser spot dependent thermal and electric conversion factors. If a dichroism is present such that $\alpha(P_1)\neq \alpha(P_2)$ for two polarization states $P_1$ and $P_2$, then the PTE produces an additional polarization-dependent photocurrent proportional to the polarization-independent contribution.

    To incorporate this PTE into the FE simulation, we introduce a thermal function:
    \begin{equation}
        T(x) = ae^{-(x - l/2)^2/(c\sigma^2)} - be^{-(x + l/2)^2/(d\sigma^2)}~,
    \end{equation}
    \noindent where $l$ is the length of the device and $\sigma$ is the Gaussian width of the laser spot. A realistic device often has asymmetries in the thermal profile, which can be adjusted by changing the relative peak height and widths ($a,b,c,d$). The results of the FE simulation with the thermal function is shown in Figure~\ref{fig:SM-thermalsim}.   
    
    \subsubsection{Linear Dichroism}
    \noindent Figure~\ref{fig:main-wavedep} of the main text shows for near-IR wavelengths that $D$ and $L_2$ are spatially dependent along the bow-tie device, consistent with a PTE driven photocurrent.  Both $D$ and $L_2$ peak near the device contacts, with opposite signs at either end. Following our discussion above, a linear dichroic absorptance near the contacts would cause the matching $D$ and $L_2$ behavior. Multiple effects could cause linear dichroism at these wavelengths: the contact/material interface provides a natural structure that may bias the absorptance of  polarization oriented parallel versus perpendicular to the contact edge, but it also creates a Schottky field~\cite{dhara_voltage-tunable_2015}, and the resulting contact field effect can modify the optical absorptance of the layered structure of \ce{PtSe2} to favor out-of-plane (P-, $E_z$) or surface-parallel (S-, $E_{xy}$) polarization as seen in other TMDCs~\cite{fan_terahertz_2020}.

    In order to determine which type of linear dichroism plays a role in position dependent measurements in Figure~\ref{fig:main-wavedep} of the main text, we perform a series of polarization dependent measurements Figure~\ref{fig:main-dichro}(a). By sending the laser through a half-wave-plate (HWP), and illuminating the bow-tie at normal incidence right at the contact/material interface we find no clear polarization dependence on the HWP angle $\psi$. For off-normal incidence, we find a $\cos(4\psi)$ dependence in the photocurrent regardless of the orientation of the bow-tie. $\psi = 0,\pi/2,\ldots$ corresponds to horizontal polarization in the lab frame, and partial p-polarization in the sample frame, while $\psi = \pi/4,3\pi/4,\ldots$ is s-polarized. For an electric field at angle of incidence (AOI) $\eta$, the in- and out-of-plane polarization components dependent on the HWP angle are:

    \begin{eqnarray}
    |E_{xy}|^2 &=& \frac{1}{4}(3+ \cos(2\eta)) + \frac{1}{2} \cos(4\psi) \sin^2(\eta)~,\\
    |E_{z}|^2 &=& \frac{1}{2}(1 - \cos(4\psi)) \sin^2(\eta)~.
    \end{eqnarray}

    \noindent The $\cos(4\psi)$ dependence on the in-plane and out-of-plane polarizations for off-normal incidence corroborates the functional form in Figure~\ref{fig:main-dichro}(a) of the main text. If we instead consider the QWP case, as in Figure~\ref{fig:main-wavedep} of the main text, the polarization components for QWP angle $\theta$ are:

    \begin{eqnarray}
    |E_{xy}|^2 &=& \frac{1}{8}(7+ \cos(2\eta)) + \frac{1}{4} \cos(4\theta) \sin^2(\eta)~,\\
    |E_{z}|^2 &=& \frac{1}{4}(1 - \cos(4\theta)) \sin^2(\eta)~.
    \end{eqnarray}

    \noindent Thus, a S vs P dichroism would also produce the $L_2$ contribution seen in Figure~\ref{fig:main-wavedep} of the main text.

    \subsubsection{Circular Dichroism}
    \noindent The MWIR spot-size of 200 $\mu$m produces a broader temperature profile in \ce{PtSe2} than in the near-IR measurements. It is therefore unlikely previously explored contact field effects are responsible for the observed MWIR helicity-dependent photocurrent due to the length scales involved~\cite{dhara_voltage-tunable_2015}. Moreover, intrinsic CPGE or CPDE helicity-dependent photocurrents are not anticipated to flip their sign as the spot is scanned across the device, whereas the measured MWIR $C$ coefficient follows the same spatial profile as $D$. Together with the linear power dependence of both coefficients (Figure~\ref{fig:main-dichro} of the main text), this suggests that the MWIR helicity-dependent photocurrent follows from a circular dichroic absorptance coupled to the PTE. A simple helicity dependent modulation of the laser power induced by the QWP is unlikely, as it would produce a similar $C/D$ ratio independent of angle of incidence. In contrast, Figure~\ref{fig:SM-wavedepnorm} shows that $C$ is suppressed for normal incidence while $D$ remains finite with a position dependent sign, indicating that the MWIR $C$ coefficient is tied to oblique-incidence sample absorptance rather than a trivial QWP power artifact.

    Following rotation of the QWP by an angle $\theta$, the circular Stokes parameter is
    \begin{equation}
        S_3 = \sin(2\theta)~.
    \end{equation}
    Therefore, a photocurrent contribution proportional to $\sin(2\theta)$ requires a net difference in absorptance for opposite helicities, i.e. $\alpha(\mathrm{LCP}) \neq \alpha(\mathrm{RCP})$. A small net crystalline canting shown in TEM and XRD combined with oblique incidence allows an extrinsic circular dichroism.

    Phenomenologically, the helicity-dependent component of the absorptance, $\alpha_h$, must be a real scalar constructed from the available optical and structural vectors. For a net tilt vector $\mathbf{t}$, surface normal $\hat{z}$, and optical spin pseudovector $\mathbf{s}_\gamma \propto i\mathbf{E}\times\mathbf{E}^*$, the lowest-order helicity-odd term allowed by symmetry is

    \begin{equation}
        \alpha_h \sim \mathbf{s}_\gamma \cdot (\mathbf{t}\times \hat{z})~.
    \end{equation}

    \noindent Since $\mathbf{s}_\gamma \propto S_3 \hat{q}$ for a plane wave, this becomes

    \begin{equation}
        \alpha_h \sim S_3\, \mathbf{t}\cdot(\hat{z}\times\hat{q})
        \sim \sin(2\theta)\,\mathbf{t}\cdot(\hat{z}\times\hat{q})~.
    \end{equation}

    \noindent Thus, a $\sin(2\theta)$ modulation of the absorptance, and therefore of the PTE photocurrent, is expected only for oblique incidence and a nonzero average tilt component perpendicular to the plane of incidence.

    \newpage

\end{document}